\documentclass[a4paper]{jpconf}
\usepackage{graphicx}

\newcommand{\tet}{\theta}
\newcommand{\az}{\varphi}

\newcommand{\oeq}{\begin{equation}}
\newcommand{\ceq}{\end{equation}}
\newcommand{\oeqn}{\begin{eqnarray}}
\newcommand{\ceqn}{\end{eqnarray}}

\renewcommand{\>}{\rangle}
\newcommand{\<}{\langle}
\renewcommand{\(}{\left(}
\renewcommand{\)}{\right)}
\renewcommand{\[}{\left[}
\renewcommand{\]}{\right]}

\newcommand{\stf}{\,\,\,}
\newcommand{\sdf}{\,\,}
\newcommand{\stb}{\!\!\!}

\newcommand{\kfi}{|\phi \>}

\newcommand{\bfi}{\<\phi |}

\newcommand{\oP}{\hat{P}}

\newcommand{\oad}{\hat{a}^\dagger}
\newcommand{\oa}{\hat{a}}

\newcommand{\oN}{\hat{N}}

\renewcommand{\d}{{\mbox d}}

\newcommand{\vr}{{\bf r}}

\newcommand{\vR}{{\bf R}}

\newcommand{\mH}{{\mathcal{H}}}

\begin{document}
\title{Quantum microscopic approach to low-energy heavy ion collisions 
}

\author{C\'edric Simenel$^{1,2}$, Aditya Wakhle$^{2}$ and Beno\^it Avez$^{1,3}$}

\address{$^{1}$ CEA, Centre de Saclay, IRFU/Service de Physique Nucl\'eaire, F-91191 Gif-sur-Yvette, France }

\address{$^{2}$ Department of Nuclear Physics, Research School of Physics and Engineering, Australian National University, Canberra, Australian Capital Territory 0200, Australia}

\address{$^{3}$Universit\'e Bordeaux 1, CNRS/IN2P3, Centre d'\'Etudes Nucl\'eaires de Bordeaux Gradignan, 
CENBG/THEO, Chemin du Solarium, BP120, 33175 Gradignan, France}

\ead{cedric.simenel@anu.edu.au}

\begin{abstract}
The Time-dependent Hartree-Fock (TDHF) theory is applied to the study of heavy ion collisions at energies around the Coulomb barrier. 
The competition between fusion and nucleon transfer mechanisms is investigated. 
For intermediate mass systems such as $^{16}$O+$^{208}$Pb, proton transfer favors fusion by reducing the Coulomb repulsion. 
A comparison with sub-barrier transfer  experimental data  shows that pairing correlations are playing an important role in enhancing proton pair transfer.
For heavier and more symmetric systems, a fusion hindrance is observed due to the dominance of the quasi-fission process. Typical quasi-fission time of few zeptoseconds are obtained. 
Actinide collisions are also investigated both within the TDHF approach and with the Ballian-V\'en\'eroni prescription for fluctuation and correlation of one-body observables. 
The possible formation of new heavy neutron-rich nuclei in actinide collisions is discussed.
\end{abstract}

\section{Introduction \label{sec:intro}}

Microscopic description of nuclear reactions at low energy where quantum effects play a significant role is an important challenge of nuclear physics. 
The interplay between nuclear structure and reaction mechanisms is crucial at energies around the Coulomb barrier generated by the competition between Coulomb and nuclear interactions.
For instance, this has been well established for fusion reactions (for a review, see, {e.g.}, Ref.~\cite{das98}).

A good understanding of these effects is crucial for the quest of superheavy elements (SHE)~\cite{hof00}. Indeed, different structures of the collision partners may change SHE production cross sections by several orders of magnitudes.
For instance, collisions of quasi-symmetric heavy systems with typical charge products $Z_1 Z_2\geq 1600$ exhibit a fusion hindrance when compared to systematics and models established for lighter systems~\cite{gag84}.
The dynamical shape evolution of the dinuclear system and the transfer between the fragments are expected to play a significant role on this fusion hindrance. 

In the case of the collision of actinide nuclei, the Coulomb repulsion is so strong that fusion cannot occur. However, a dinuclear system may survive up to few zeptoseconds, forming the heaviest nuclear systems available on Earth. Multi-nucleon transfer between actinides may provide an alternate way to produce more neutron rich heavy and super-heavy elements. The complex dynamics of such systems with up to $\sim500$ nucleons in interaction is very challenging and has been investigated with different theoretical approaches recently~\cite{ada05,fen09,zag06,zag12,mar02,tia08,zha09,gol09,ked10}.

The time-dependent Hartree-Fock (TDHF) theory~\cite{dir30} presents the advantage of treating both structure  and reaction dynamics within the same formalism. 
The TDHF theory provides a self-consistent mean-field description of the many-body dynamics. 
Early TDHF calculations in nuclear physics used various symmetries and 
simplified Skyrme interactions~\cite{sky56} to reduce computational 
times~\cite{eng75,bon76,neg82}.
Recent increase of computational power allowed 
realistic TDHF calculations in 3 dimensions \cite{kim97,mar05,nak05,uma05,seb09} 
with full Skyrme energy density functional 
(EDF) including spin-orbit terms \cite{sky56,cha98}. 
Several examples of recent applications of TDHF to nuclear dynamics were presented at the NN2012 conference~\cite{ita12,iwa12b,obe12b,mot12}.

In Section~\ref{sec:TDHF}, we recall the TDHF formalism and provide some details on the calculations.
In Section~\ref{sec:fusion}, we study  the interplay between fusion and transfer in the $^{16}$O+$^{208}$Pb system. Section~\ref{sec:fusion} is dedicated to the description of sub-barrier transfer in this reaction. 
In Section~\ref{sec:QF}, we investigate the fusion hindrance of heavy quasi-symmetric systems and the quasi-fission process. Finally, we study the collision dynamics of actinide nuclei
in Section~\ref{sec:actinides} before to conclude.

\section{The Time-Dependent Hartree-Fock approach}
\label{sec:TDHF}

In nuclear physics, the TDHF theory is applied with a Skyrme EDF modeling nuclear interactions between nucleons~\cite{sky56}. The EDF is the only phenomenological ingredient of the model, as it has been adjusted on nuclear structure properties like infinite nuclear matter and radii and masses of few doubly magic nuclei~\cite{cha98}. The main approximation of the theory is to constrain the many-body wave function to be an antisymmetrized independent particle state at any time. The latter ensures an exact treatment of the Pauli principle during time evolution. Though TDHF does not include two-body collision term, it is expected to treat correctly one-body dissipation which is known to drive low energy reaction mechanisms as Pauli blocking prevents nucleon-nucleon collisions.

The main advantage of TDHF is that it treats static
properties {\it and} dynamics of nuclei within the same
formalism and the same EDF. The initial state is obtained
through static HF calculations which 
give a good approximation of nuclear binding energies and
deformations~\cite{vau72,ben03}. 
Another important advantage of TDHF for near-barrier reaction studies 
is that it contains all  types of couplings between the relative
motion and internal degrees of freedom.
However, TDHF gives only classical trajectories for the time-evolution  and expectation values of one-body observables. In particular, it does not include tunneling of the many-body wave function and may underestimate fluctuations of one-body observables~\cite{das79,bal81}.
The latter fluctuations can be estimated in the time-dependent RPA limit using a prescription proposed by Balian and V\'en\'eroni~\cite{bal84}.

Inclusion of pairing correlations responsible for superfluidity in nuclei have been done  recently to study  vibrations in nuclei~\cite{ave08,ass09,eba10,ste11}. However, realistic applications to heavy ion collisions are not yet achieved and are beyond the scope of this work.

\subsection{Formalism}

The TDHF equation can be written as a Liouville-Von Neumann equation 
\begin{equation}
i\hbar \frac{\partial}{\partial t} \rho = \left[h[\rho],\rho\right]
\label{eq:tdhf}
\end{equation}
where $\rho$ is the one body density matrix associated to the total independent particle state with elements 
\begin{equation}
\rho(\mathbf{r} sq, \mathbf{r'}s'q') = \sum_{i=1}^{A_1+A_2} \sdf  \varphi_i(\mathbf{r} sq)\sdf \varphi_i^*(\mathbf{r'}s'q').
\end{equation}
The sum runs over all occupied single particle wave functions $\varphi_i$ and $\mathbf{r}$, $s$ and $q$ denote the position, spin and isospin of the nucleon, respectively.
The Hartree-Fock single particle Hamiltonian $h[\rho]$ is related to the EDF, noted $E[\rho]$, by its first derivative
\begin{equation}
h[\rho](\mathbf{r} sq, \mathbf{r'}s'q') = \frac{\delta E[\rho]}{\delta \rho(\mathbf{r'} s'q', \mathbf{r} sq)}.
\end{equation}

\subsection{Practical aspects}

A TDHF calculation of two colliding nuclei is performed assuming that 
the two collision partners are initially at a distance
$D_0$ in their HF ground state. This distance has to be large enough 
to account for  Coulomb excitation in the entrance channel 
(polarization, vibration, rotation...).
Typical initial distances are within the range $D_0\sim30-50$~fm.
We assume that before the initial time,
the nuclei followed a Rutherford trajectory.
This assumption determines the 
initial velocities in the center of mass frame. 
The collision partners are put into motion using a 
Galilean boost~\cite{tho62}  which is applied on each nucleus 
at the first iteration.

We use the  {\textsc{tdhf3d}} code built by P. Bonche and coworkers with the SLy4$d$
 Skyrme parametrization~\cite{kim97}.
 This code has a plane of symmetry (the collision plane).
It uses the Skyrme energy functional expressed in Eq.~(A.2) of Ref.~\cite{bon87} 
where the tensor coupling between spin and gradient has been neglected.
The lattice spacing is $\Delta x=0.8$~fm and the time step is $\Delta t=1.5\times10^{-24}$~s.
More details on modern TDHF calculations to nuclear dynamics can be found in Refs.~\cite{sim10a,sim12b}.

\section{Fusion with medium mass systems\label{sec:fusion}}

A natural application of the TDHF approach in nuclear physics is to study fusion reactions at and above the barrier.  
Fusion occurs by transferring relative motion into internal excitation via one-body mechanisms well treated by the TDHF approach.
As a result, 
early TDHF codes have been successfully applied to describe above-barrier fusion reaction in light systems~\cite{bon78}.
In this section, we investigate the fusion process in the $^{16}$O+$^{208}$Pb system.

A reference nucleus-nucleus potential could be obtained in the frozen approximation with HF (or HFB) densities~\cite{den02}, where the collision partners are assumed to keep their ground-state density during the approach. 
The frozen potential can be computed with the same Skyrme EDF as in the TDHF calculations by translating the nuclei in their HF state. 
A comparison between TDHF and frozen fusion barriers allows to identify the role of dynamical effects, which are included in TDHF but absent in the frozen approach. 

Writing the HF energy $E[\rho]$ as an integral of an energy density $\mH[\rho(\vr)]$, i.e., 
\oeq
E[\rho]=\int \d\vr \sdf \mH[\rho(\vr)],
\ceq
we get the expression for the frozen potential 
\oeq
V(\vR)=\int \d\vr \sdf \mH[\rho_1(\vr)+\rho_2(\vr-\vR)] - E[\rho_1] -E[\rho_2],
\label{eq:frozen}
\ceq
where $\vR$ is the distance between the centers of mass of the nuclei, and $\rho_{1,2}$ are the densities of their 
HF ground-state. 
Eq.~(\ref{eq:frozen}) neglects the Pauli principle between the nucleons of one nucleus and the nucleons of the other one. 
As a result, it is valid only for small overlaps between the two collision partners. 
However, for light and intermediate mass systems, the barrier radius is large enough to justify this approximation. 

\begin{figure}
\begin{center}
\includegraphics[width=10cm]{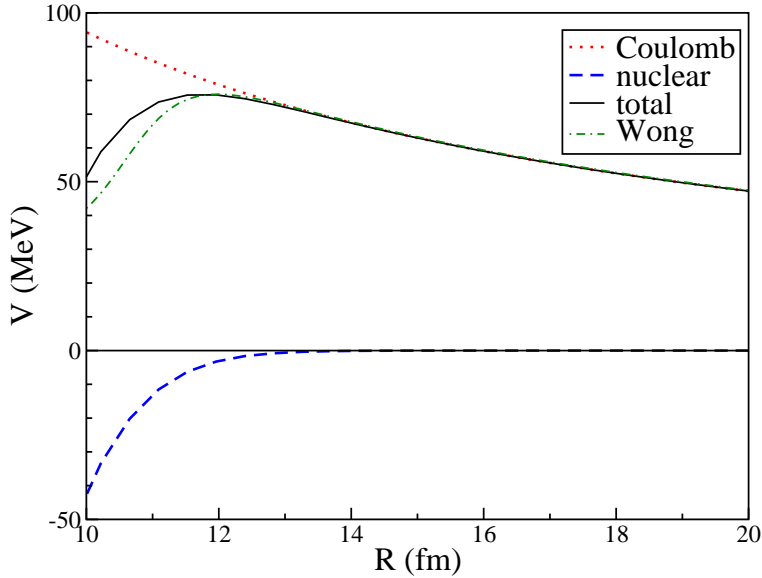}
\caption{Nuclear (dashed blue line) and Coulomb (dotted red line) contributions of the nucleus-nucleus potential (solid line) obtained with the HF-frozen approximation in the $^{16}$O+$^{208}$Pb system. 
The Wong potential~\cite{won73} is shown in green dot-dashed line. The latter is obtained with a potential depth  $V_0=70$~MeV, a potential diffuseness $a=0.48$~fm, and nuclear radii $R_i=1.25A_i^{1/3}$~fm. 
\label{fig:frozen}}
\end{center}
\end{figure}

\begin{figure}
\begin{center}
\includegraphics[width=10cm]{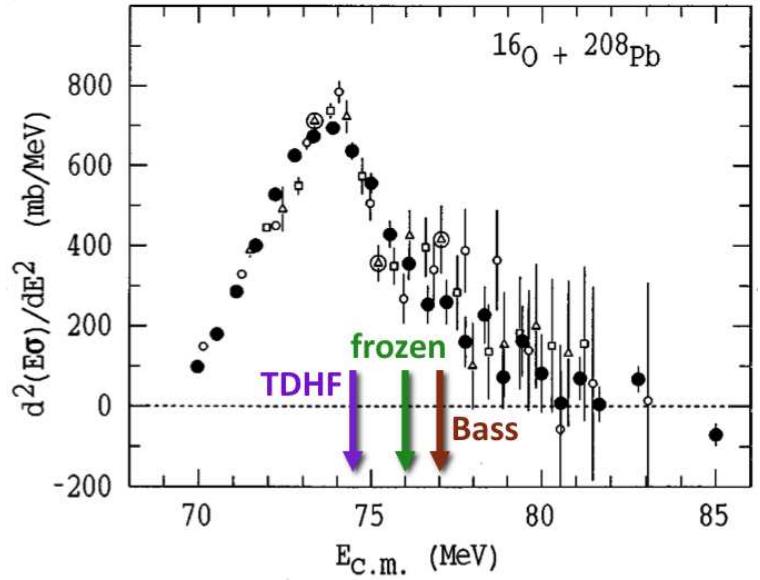}
\caption{Experimental fusion barrier distribution of the $^{16}$O+$^{208}$Pb system from Ref.~\cite{mor99}. The HF-frozen, Bass and TDHF barriers are indicated by  arrows. 
\label{fig:dist_barr}}
\end{center}
\end{figure}

The HF-frozen potential for the $^{16}$O+$^{208}$Pb system is shown in Fig.~\ref{fig:frozen}.
The potential obtained from the Wong formula~\cite{won73} is also presented for comparison. 
The height of the barrier obtained with the HF-frozen approximation is $V_B^{frozen}\simeq76.0$~MeV at $R_B^{frozen}\simeq11.8$~fm. Note that the same barrier height has been obtained in Refs.~\cite{was08,guo12} with a similar implementation of the HF-frozen approximation. 
This value is close to the barrier obtained with the Wong formula~\cite{won73}, $V_B^{Wong}\simeq75.9$~MeV, while it is 1 MeV smaller than the Bass barrier~\cite{bas77}, $V_B^{Bass}\simeq77.0$~MeV at $R_B^{Bass}\simeq11.4$~fm.
A comparison with the experimental barrier distribution shown in Fig.~\ref{fig:dist_barr} indicates that 
all these barriers overestimate the experimental value $V_B^{exp.}\sim74.5$~MeV obtained from the centroid of the barrier distribution.

We now investigate the possible role of dynamical effects on the fusion barrier with TDHF calculations. 
The TDHF fusion barrier is  identified as the capture threshold for central collisions, above which a compound system is formed and below which two fragments are emitted. 
Due to the finite time of the TDHF evolutions, one has to define a maximum computational time ($\sim10^3$~fm/$c$ in the present case) above which the final configuration (i.e., one compound system or two fragments) is assumed to be reached. 

\begin{figure}
\begin{center}
\includegraphics[width=10cm]{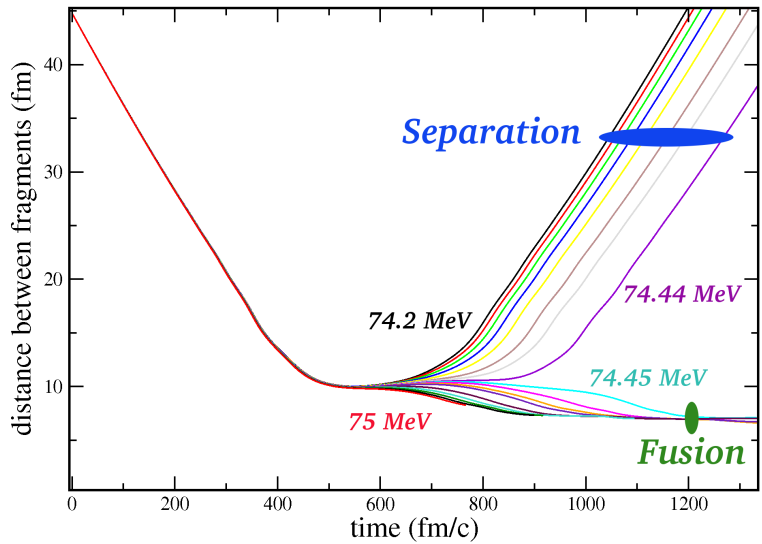}
\caption{Relative distance between fragments as a function of time for head-on 
$^{16}$O+$^{208}$Pb collisions.\label{fig:distances}}
\end{center}
\end{figure}

Figure~\ref{fig:distances} shows the evolution of the relative distance between fragment centers of mass in central $^{16}$O+$^{208}$Pb collisions at different energies around the barrier. 
We identify two sets of trajectories associated to fusion for $E\ge74.45$~MeV and to re-separation of the fragments for $E\le74.44$~MeV.
The resulting fusion barrier predicted by these TDHF calculations is, then, $V_B^{TDHF}=74.445\pm0.005$~MeV. 
As a result, the dynamical effects included in TDHF calculations lower the barrier by $\sim1.5$~MeV for this system as compared to the HF-frozen approximation. 
Other methods based on a macroscopic reduction of the mean-field dynamics, namely the dissipative-dynamics TDHF \cite{was08} and the density-constrained TDHF \cite{uma09b},
also find similar results with an energy dependence to the
barrier heights ranging from 74.5 at low energies to 76 MeV
at higher energies where the frozen approach is more valid. 

\begin{figure}
\begin{center}
\includegraphics[width=15cm]{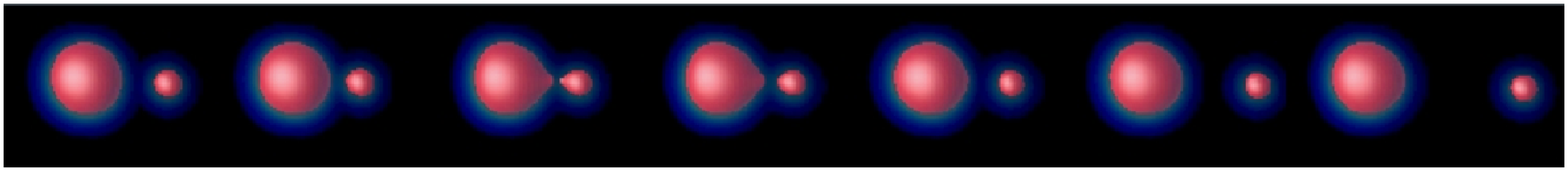}
\includegraphics[width=15cm]{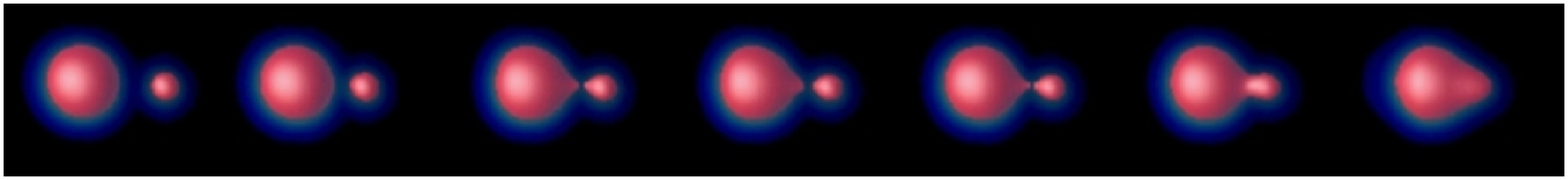}
\end{center}
\caption{(top) Density evolution for the reaction $^{16}$O+$^{208}$Pb corresponding to a head-on collision at a center of mass energy $E_{c.m.}=74.44$~MeV (just below the fusion barrier). The red surfaces correspond to an iso-density at half the saturation density ($\rho_0/2=0.08$~fm$^{-3}$). Each figure is separated by a time step of 135~fm/c. Time runs from left to right. (bottom) Same at $E_{c.m.}=74.45$~MeV, i.e., just above the fusion threshold.\label{fig:dens}}
\end{figure}

The density evolutions close to the fusion threshold, i.e., at $E_{c.m.}=74.44$ and 74.45~MeV, are plotted in Fig.~\ref{fig:dens}.
In both cases, a ''di-nuclear'' system is formed during a relatively long time ($\sim500$~fm/c) before either re-separation or fusion.
Nucleon transfer is expected to occur in this di-nuclear system.
An evidence of this transfer is given by integrating the proton and nucleon densities at the final time when the two fragments re-separate at $E_{c.m.}=74.44$~MeV. As a result, the outgoing channel is, in average, $^{14}$C+$^{210}$Po, showing the importance of proton transfer from the light to the heavy partner, in good agreement with experiment~\cite{vid77,eve11}.
This proton transfer effectively lowers the barrier by decreasing $Z_1Z_2$ and, then, the Coulomb repulsion. 
As a result, the coupling to transfer channels increase the fusion probability for this system.
Transfer reactions in the $^{16}$O+$^{208}$Pb system are discussed in more details in the next section. 
Note  that low-lying collective vibrations are also known to affect the fusion barrier distribution~\cite{mor99}.

\section{Transfer reactions \label{sec:transfer}}

Transfer reactions in heavy-ion collisions have been investigated within the TDHF framework recently~\cite{uma08a,sim08,was09b,sim10a,sim10b,yil11}.
In fact, the previous study showed the importance of the interplay between fusion and transfer reactions in the $^{16}$O+$^{208}$Pb system around the barrier. 
For instance, just below the barrier, the light fragment in the exit channel (see top of Fig.~\ref{fig:dens}) is, in average, a $^{14}$C nucleus, indicating that the transfer of two protons is the dominant channel at this energy.
We now discuss this sub-barrier transfer reaction in more details.

\begin{figure}
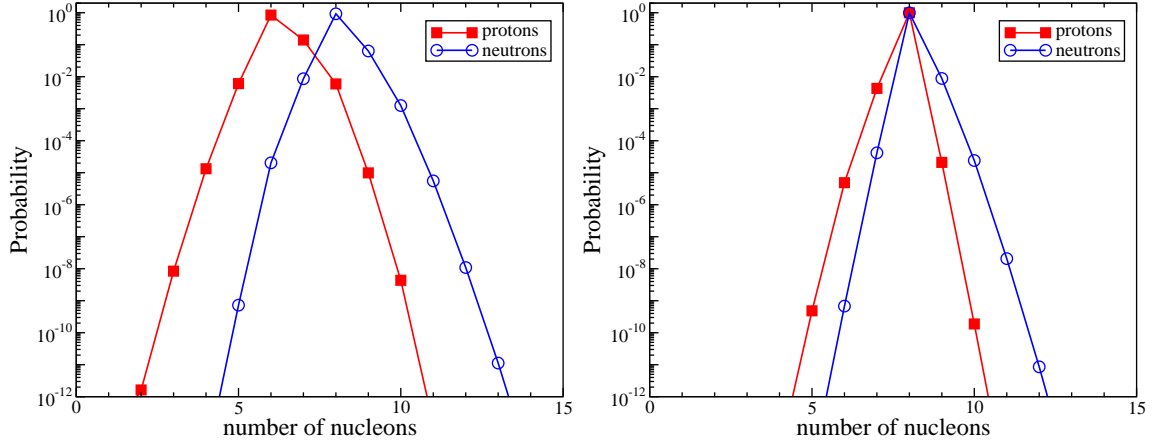

\begin{center}
\includegraphics[width=7.5cm]{proba_B.eps} 
\includegraphics[width=7.5cm]{proba_subB.eps} 
\caption{Neutron (circles) and proton (squares) number probability distributions 
of the lightest fragment in the exit channel of a head-on 
$^{16}$O+$^{208}$Pb collision at $E_{c.m.}=74.44$~MeV (left) and $65$~MeV (right). Adapted from Ref.~\cite{sim10b}.}
\label{fig:proba}
\end{center}
\end{figure}

In the previous section, a signature for transfer mechanism was obtained from the change of average numbers of nucleons in the fragments. 
Using the quantum nature of the TDHF theory, it is possible to compute the probability for a specific outgoing channel, i.e., the probability to find a given number of neutrons and protons in the fragments~\cite{koo77,sim10b}.  
This is possible, e.g., thanks to a particle number projection technique applied on the outgoing fragments~\cite{sim10b}.

Let us consider a final time when the two fragments are located on both side of the $x=0$ plane and define a particle number projector onto $N$ protons or neutrons in the $x>0$ region
\oeq
\oP_R(N)=\frac{1}{2\pi}\int_0^{2\pi} \stb \d \tet \stf e^{i\tet(\oN_R-N)},
\label{eq:projector}
\ceq
where
\oeq
\oN_R = \sum_{s} \sdf \int \stb \d \vr \stf \oad(\vr s) \sdf \oa(\vr s) 
\sdf \Theta(x)
\label{eq:NG}
\ceq
counts the number of particles in the $x>0$ region ($\Theta(x)=1$ if $x>0$ and 0 elsewhere).
Isospin is omitted to simplify the notation. 
The projector defined in Eq.~(\ref{eq:projector}) can be used to compute the probability to find $N$ nucleons in $x>0$ in the final state $\kfi$,
\oeq
\left|\oP_R(N)\kfi\right|^2=\<\phi|\oP_R(N) |\phi\>=\frac{1}{2\pi}\int_0^{2\pi} \stb \d \tet \stf e^{-i\tet{N}}\bfi\phi_R(\tet)\>,
\label{eq:proba}
\ceq
where $|\phi_R(\tet)\>=e^{{i\tet\oN_R}}\kfi$.
Note that $|\phi_R(\tet)\>$ is an independent particle state and, then, 
the last term in Eq.~(\ref{eq:proba}) is  the determinant of the matrix of the occupied single particle state overlaps~\cite{sim10b}:
\oeq
\bfi\phi_R(\tet)\>=\det (F)
\ceq
with
\oeq
F_{ij}= \sum_{s} \int \stb\d \vr \sdf{\az_i^s}^*(\vr) {\az_j^{s}}(\vr) e^{i\tet\Theta(x)}.
\ceq
The integral in Eq.~(\ref{eq:proba}) is discretized using $\tet_n=2\pi{n}/M$ with the integer $n=1\cdots{M}$.
Choosing $M=300$ ensures numerical convergence for the $^{16}$O+$^{208}$Pb system. 
Fig.~\ref{fig:proba} shows the resulting transfer probabilities at $E_{c.m.}=74.44$~MeV (left) and at $E_{c.m.}=65$~MeV (right).
In agreement with the results presented in section~\ref{sec:fusion}, the most probable channel is $Z=6$ and $N=8$ at the barrier. However, lowering the energy reduces the transfer probabilities and the main channel is $Z=N=8$ well below the barrier, corresponding to inelastic and elastic channels. 

\begin{figure}
\begin{center}
\includegraphics[width=10cm]{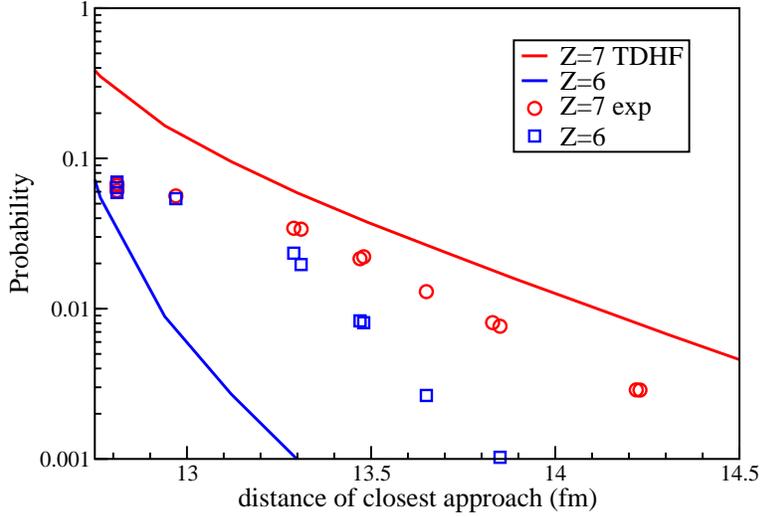} 
\caption{Proton number probability  
as function of the distance of closest approach  in the small outgoing fragment of the $^{16}$O+$^{208}$Pb reaction. TDHF results are shown with lines. Experimental data (open symbols) are taken from Ref.~\cite{eve11}.}
\label{fig:proba_TDHF+exp}
\end{center}
\end{figure}

To compare with experimental  data  on the $^{16}$O+$^{208}$Pb reaction, we plot  in Fig.~\ref{fig:proba_TDHF+exp}  the transfer probabilities from TDHF as a function of the distance of closest approach $R_{min}$ between the collision partners~\cite{cor09}, assuming a Rutherford trajectory~\cite{bro91}:
\oeq
R_{min}={Z_1Z_2e^2}[1+\mbox{cosec}(\theta_{c.m.}/2)]/{2E_{c.m.}}
\label{eq:R_min}
\ceq
where $\theta_{c.m.}$ is the center of mass scattering angle.
Recent data from Ref.~\cite{eve11} are shown in Fig.~\ref{fig:proba_TDHF+exp} for sub-barrier one- and two-proton transfer channels.
We see that TDHF overestimates the one-proton transfer probabilities and underestimates the strength of the two-proton transfer channel.
This discrepancy is interpreted as an effect of pairing interactions~\cite{sim10b,eve11}.
Indeed, paired nucleons are expected to be transferred as a pair, increasing (resp. decreasing) the two-nucleon (single-nucleon) transfer probability. 
For $R_{min}>13$~fm, however, the TDHF calculations reproduces reasonably well the sum of one and two-proton transfer channels. For $R_{min}<13$~fm,  sub-barrier fusion, not included in the TDHF calculations, reduces transfer probabilities~\cite{sim10b,eve11}.

These studies emphasize the role of pairing interactions in heavy-ion collisions.
The recent inclusion of pairing interactions in 3-dimensional microscopic codes~\cite{ass09,eba10,ste11} gives hope in our ability to describe such data more realistically  in a near future. 

\section{Fusion and quasi-fission in heavy systems\label{sec:QF}}

To a reasonably good approximation, the fusion barrier for light and intermediate  mass systems is determined by the frozen barrier. 
This approximation fails, however, for heavy systems with typical charge products $Z_1 Z_2\geq 1600$ which are known to exhibit fusion hindrance~\cite{gag84}. 
Above this threshold, an extra-push energy is usually needed for the system to fuse~\cite{swi82}.
In fact, at the energy of the frozen barrier, heavy systems are more likely to encounter quasi-fission, i.e., a re-separation in two fragments after a possible mass exchange. 

We now investigate the reaction mechanism in heavy systems with possible fusion hindrance.
We first illustrate the fusion hindrance with TDHF calculations of fusion thresholds.  
Then, we investigate the quasi-fission process. 

\subsection{TDHF calculations of fusion hindrance}

\begin{figure}
\begin{center}
\includegraphics[width=10cm]{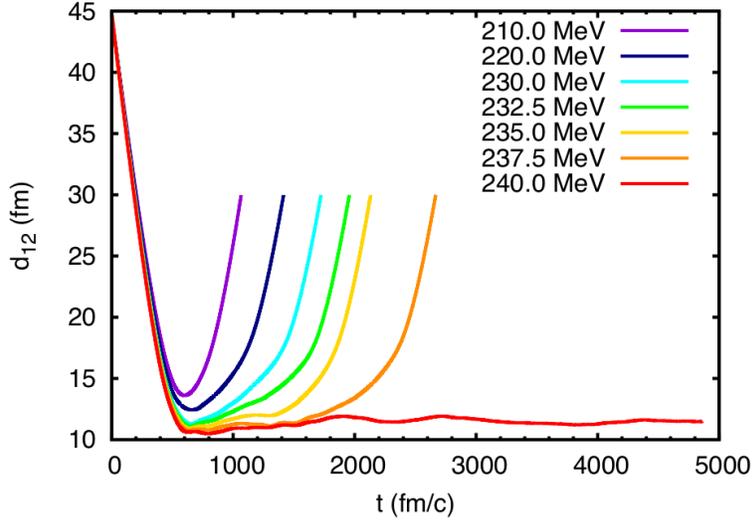} 
\caption{Distance between the centers of mass of the fragments as a function of time in head-on 
$^{90}$Zr+$^{124}$Sn collisions for different center of mass energies. }
\label{fig:ZrSn}
\end{center}
\end{figure}

Let us first consider the $^{90}$Zr+$^{124}$Sn system which has a charge product $Z_1Z_2=2000$, and, then, 
is expected to exhibit a fusion hindrance. 
Indeed, the proximity model~\cite{blo77} predicts a barrier for this system $V^{prox.}\simeq215$~MeV, while TDHF calculations predicts 
that the system encounters a fast re-separation at this energy, as shown in Fig.~\ref{fig:ZrSn}~\cite{ave09}. 
In the same figure, we observe a long contact time at $E_{c.m.}=240$~MeV, which is interpreted as a capture trajectory leading to fusion. 
The additional energy needed to fuse is then $\sim25$~MeV, which is higher than the  extra-push model ~\cite{swi82} prediction $E^{X-push}\simeq14.8$~MeV.

\begin{figure}
\begin{center}
\includegraphics[width=10cm]{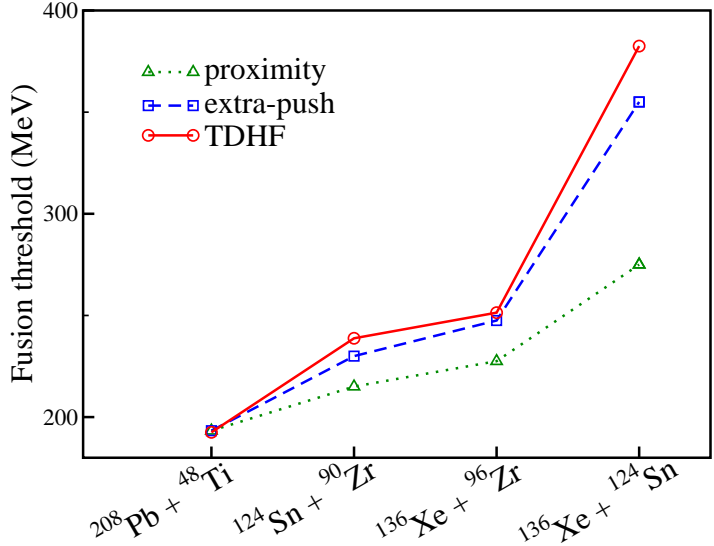} 
\caption{TDHF fusion thresholds for several heavy systems are compared with the proximity barrier~\cite{blo77} and with results from the extra-push model~\cite{swi82}.
}
\label{fig:xpush}
\end{center}
\end{figure}

TDHF calculations of fusion hindrance have been performed for several systems, as shown in Fig.~\ref{fig:xpush}, where the TDHF fusion thresholds are compared with the interaction barriers predicted by the proximity model~\cite{blo77} and with the extra-push  model~\cite{swi82}.
We observe an increase of the fusion threshold with TDHF as compared to the proximity model which assumes frozen reactants. 
This indicates that dynamical effects are playing an important role in the reaction by hindering  fusion. 
The order of magnitude of the additional energy needed to fuse is similar to the one predicted with the extra-push model. 

\subsection{The quasi-fission process\label{sec:QFTDHF}}

\begin{figure}
\begin{center}
\includegraphics[width=15cm]{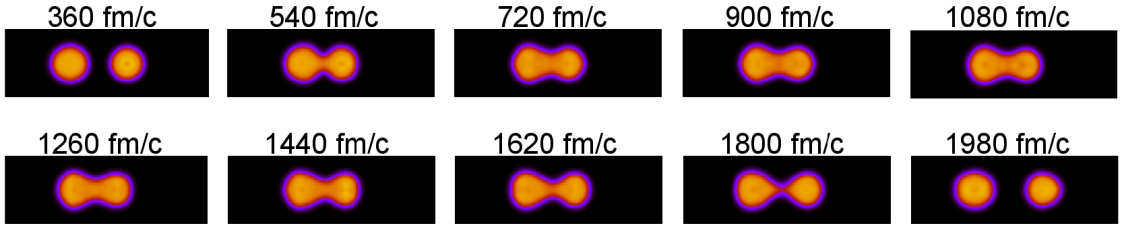} 
\caption{Density profile in the $^{90}$Zr+$^{124}$Sn  head-on collision at $E_{c.m.}=235$~MeV. From Ref.~\cite{ave09}.}
\label{fig:ZrSn_dens}
\end{center}
\end{figure}

The quasi-fission mechanism is a fast re-separation of the fragments, with usually a partial  equilibration of their mass. Typical quasi-fission times are of the order of few zepto-seconds~\cite{tok85,rie11,sim12b}.
Quasi-fission becomes dominant in heavy systems and is mostly responsible for the fusion hindrance discussed in the previous section.
As an example, Fig.~\ref{fig:ZrSn_dens} shows the density profiles for the $^{90}$Zr+$^{124}$Sn head-on collision at $E_{c.m.}=235$~MeV.
We observe that the two fragments are in contact during  $\sim5$~zs. 
A dinuclear system is then formed before re-separation in two fission-like fragments.

\begin{figure}
\begin{center}
\includegraphics[width=10cm]{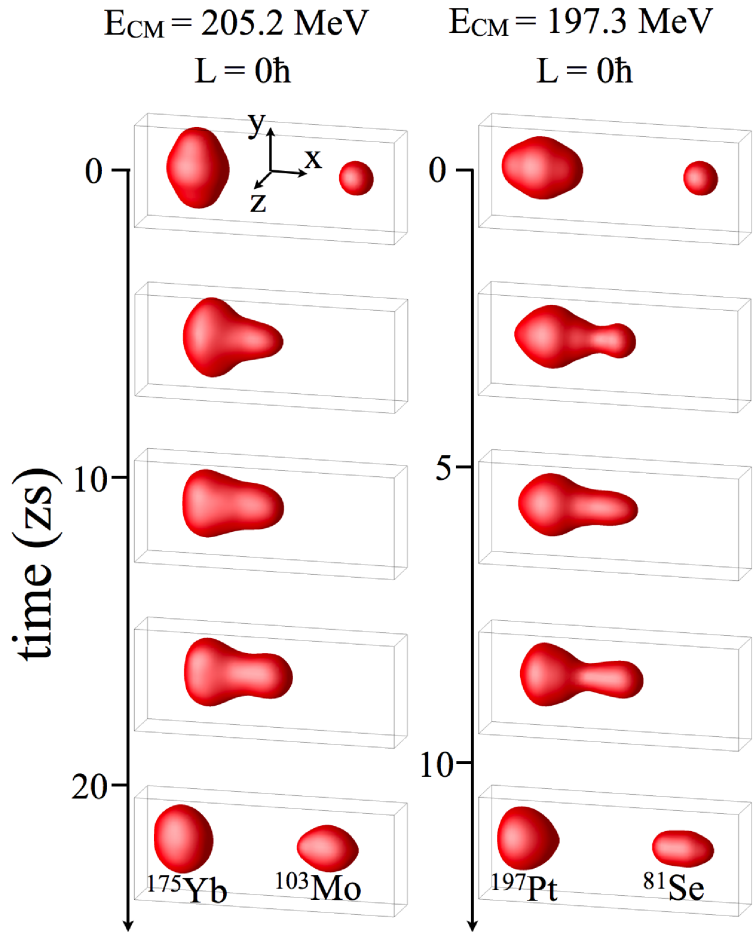}
\caption{Snapshots of the TDHF isodensity at half the saturation density in the $^{40}$Ca+$^{238}$U system for different initial orientations and $E_{c.m.}$.}
\label{fig:densQF}
\end{center}
\end{figure}

Extensive calculations are ongoing on the $^{40}$Ca+$^{238}$U system to compare with recent measurements  performed at the Australian National University~\cite{wak12b}. 
Examples of density evolutions are shown in Fig.~\ref{fig:densQF} for two different initial conditions.
In both cases, a quasi-fission process is obtained. 
In particular, we observe an important multi-nucleon transfer from the heavy fragment toward the light one. 

\begin{figure}
\begin{center}
\includegraphics[width=10cm]{mass.eps}
\hspace{0.5cm}
\includegraphics[width=10cm]{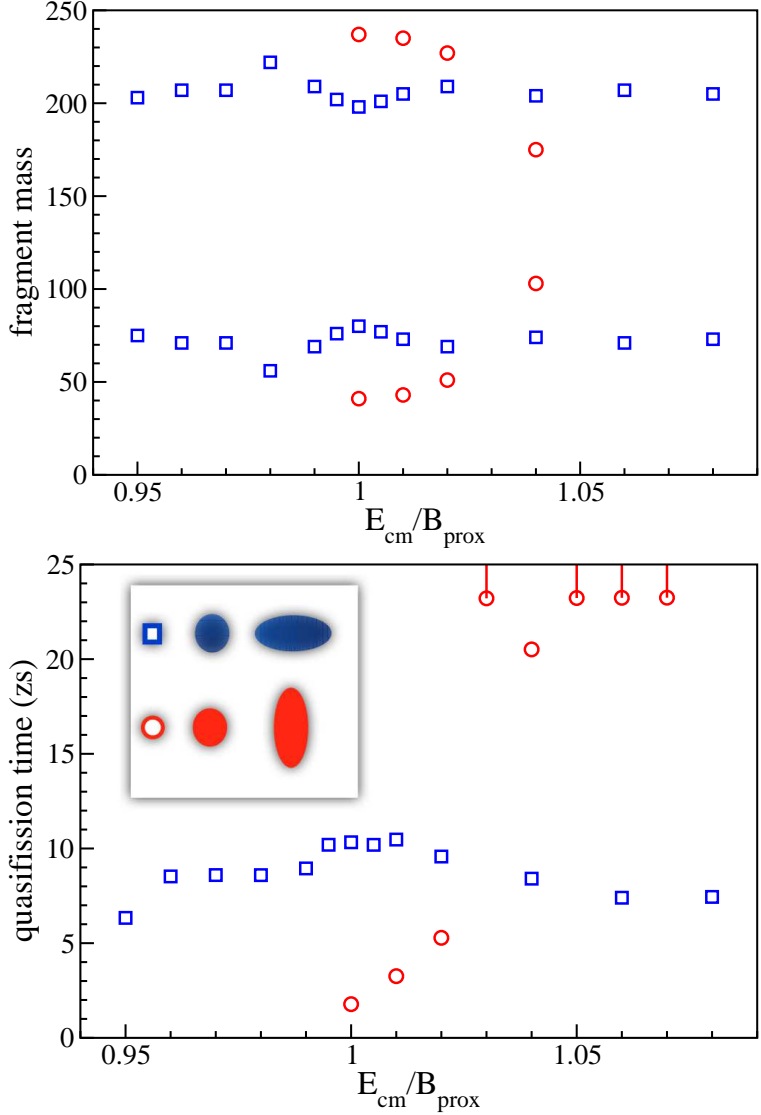}
\caption{TDHF calculations of the mass of the fragments (top) and of the quasi-fission time (bottom) in $^{40}$Ca+$^{238}$U central collisions as a function of the center of mass energy (divided by the proximity barrier~\cite{blo77}). For quasi-fission times larger than 23~zs, only a lower limit is given.  Two different orientations of the $^{238}$U are considered (see inset). }
\label{fig:QFtime}
\end{center}
\end{figure}

Fig.~\ref{fig:QFtime} presents final fragment masses (top) and quasi-fission times (bottom) for two different orientations of the $^{238}$U. 
Comparing these two figures, we observe that the mass equilibration (i.e., the formation of two fragments with symmetric masses) is not complete and varies with the life-time of the dinuclear system, i.e., the longer the contact, the larger the mass transfer. 
We also see that all the calculations with the $^{238}$U deformation axis aligned with the collision axis lead to a quasi-fission with partial mass equilibration and quasi-fission times smaller than 10~zs. 
Shell effects may affect the final outcome of the reaction by favouring the production of fragments in the $^{208}$Pb region.
In particular, this orientation never leads to fusion, while the other orientation produces long contact times above the barrier which may be associated to fusion. 
Calculations of non-central $^{40}$Ca+$^{238}$U collisions are ongoing in order to compare with experimental data.

\section{Actinide collisions \label{sec:actinides}}

Collisions of actinides form, during few zs, the heaviest nuclear systems available on Earth. 
In one hand, such systems are interesting to study the stability of the QED vacuum under strong electric fields~\cite{rei81,ack08,gol09}. 
In the other hand, they might be used to form neutron-rich heavy and super-heavy elements via multi-nucleon transfer reactions~\cite{zag06,zag12,ked10}.

\begin{figure}
\begin{center}
\includegraphics[width=10cm]{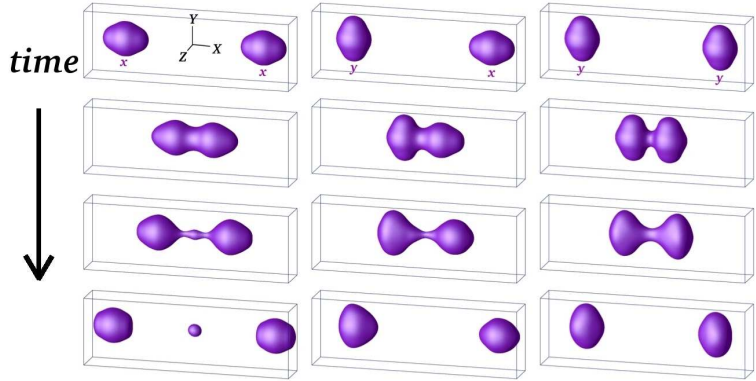}
\caption{Snapshots of the isodensity at half the saturation density in $^{238}$U+$^{238}$U central collisions at $E_{c.m.}=900$~MeV from TDHF calculations. Snapshots are given at times $t=0$, 1.5, 2.7, and $4.2$~zs from top to bottom. From Ref.~\cite{gol09}.}
\label{fig:density}
\end{center}
\end{figure}

Actinide collisions have been studied with the TDHF approach~\cite{cus80,gol09,ked10,sim11b}.
Fig.~\ref{fig:density} shows  density evolutions in $^{238}$U+$^{238}$U central collisions at $E_{c.m.}=900$~MeV for different initial conditions. 
We see that the initial orientation of the nuclei plays a crucial role on the reaction mechanism~\cite{gol09}, with the production of a third fragment in the tip-tip collision (left), or net mass transfer in the tip-side configuration (middle).
In the latter case, $\sim6$~protons and $\sim11$~neutrons, in average, are transferred from the right to the left nucleus, corresponding to the formation of a $^{255}$Cf primary fragment. 

\begin{figure*}
\begin{center}
\includegraphics[width=15cm]{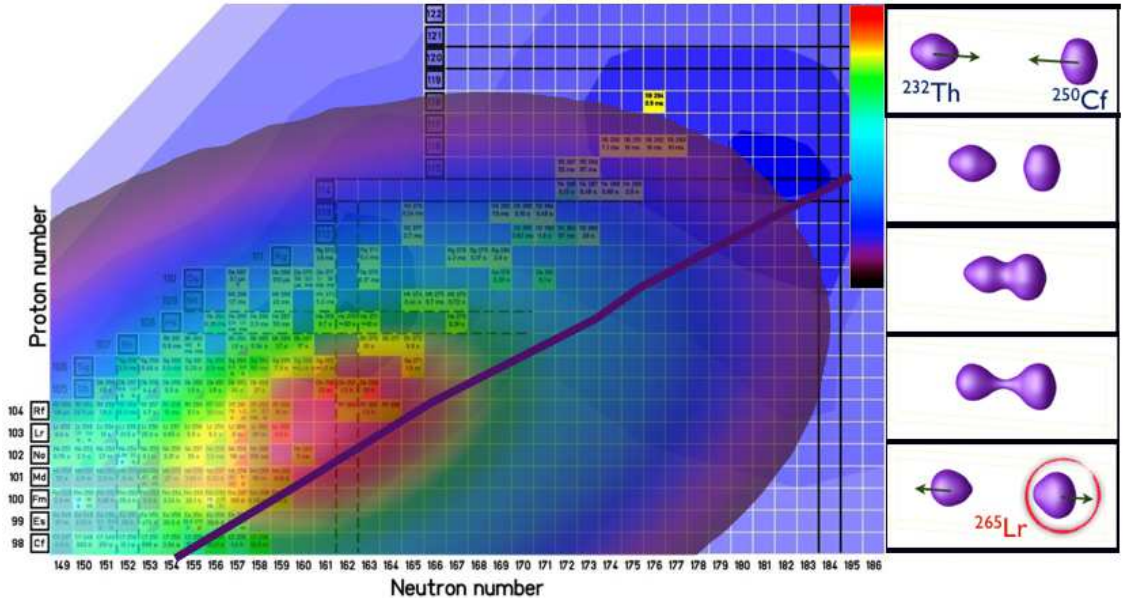}
\caption{(right) Snapshots of the isodensity at half the saturation density in $^{232}$Th+$^{250}$Cf central collisions at $E_{c.m.}=916$~MeV. (left) Gaussian distributions of $N$ and $Z$ with widths and correlations computed with the BV prescription (linear color scale). The solid line represents the predicted $\beta-$stability line.}
\label{fig:dist}
\end{center}
\end{figure*}

Similar calculations have been performed on the $^{232}$Th+$^{250}$Cf system~\cite{ked10}.
An example of density evolution is shown in the right panel of Fig.~\ref{fig:dist}.
In this case, we observe a net transfer of nucleons from the tip of the $^{232}$Th to the side of the $^{250}$Cf, corresponding to an inverse quasi-fission process, i.e., the exit channel is more mass-asymmetric than the entrance channel. Indeed, in this case, a $^{265}$Lr fragment is formed in the exit channel.
It is worth mentioning that these calculations predict inverse quasi-fission for this specific orientation only, i.e., when the tip of the lighter actinide is in contact with the side of the heavier one. Indeed, the other orientations induce ''standard'' quasi-fission~\cite{ked10}. 
Note also that another inverse quasi-fission mechanism is predicted in actinide collisions due to shell effects in the $^{208}$Pb region~\cite{vol78,zag06,zag12}. 

In the previous example, the $^{265}$Lr heavy fragment indicates the average $N$ and $Z$ of a distribution. 
The fluctuations and correlations of these distributions have been computed with the Balian-V\'en\'eroni prescription~\cite{bal84} using the \textsc{tdhf3d} code~\cite{sim11,sim11b}.
Fig.~\ref{fig:dist}(left) shows the resulting probabilities assuming Gaussian distributions of the form 
\oeq
P(z,n)=\(2\pi\sigma_N\sigma_Z\sqrt{1-\rho^2}\)^{-1} \exp\[ -\frac{1}{1-\rho^2} \( \frac{n^2}{\sigma_N^2} + \frac{z^2}{\sigma_Z^2} -\frac{2\rho nz}{\sigma_N\sigma_Z} \) \],
\label{eq:Gauss}
\ceq
where $\sigma_{N,Z}$ are the standard deviations and  $0\le\rho<1$ quantifies the correlations between the $N$ and $Z$ distributions~\cite{sim12b}.
We see that many new $\beta-$stable and neutron-rich heavy nuclei could be produced if their excitation energy is low enough to allow their survival against fission. 
This inverse quasi-fission process needs further investigations, 
in particular to predict realistic production cross-sections.

\section{Conclusions and perspectives}

The TDHF theory has been applied to the study of heavy ion collisions at energies around the Coulomb barrier. 
Its time-dependent nature allows to investigate dynamical effects responsible for the modification of the fusion thresholds. 
In particular, the coupling between the relative motion and proton transfer  in $^{16}$O+$^{208}$ decreases the barrier by reducing the Coulomb repulsion. 

Particle transfer probabilities are predicted using a particle-number projection technique. 
A comparison with experimental data shows the importance of pairing correlations on transfer.
The latter could be studied with new time-dependent Hartree-Fock-Bogoliubov codes. 
In particular, one could answer the question on the origin of these pairing correlations: Are they present in the ground-states or are they generated dynamically during the collision? 

For heavy and (quasi-)symmetric systems, a fusion hindrance is observed due to the quasi-fission process. 
The possibility to study the quasi-fission mechanism with a fully microscopic quantum approach such as the TDHF theory is promising. 
It will help to understand the strong fusion hindrance  in quasi-symmetric heavy systems and may provide a guidance for new fusion experiments with exotic beams. 

Actinide collisions have been investigated both within the TDHF approach and with the Ballian-V\'en\'eroni prescription. 
A new inverse quasi-fission mechanism associated to specific orientations was found.
This mechanism might help to produce new neutron-rich heavy nuclei.
A systematic investigation of this effect is mandatory to help the design of future experimental equipments dedicated to the study of fragments produced in actinide collisions.

\section*{Acknowledgements}

D. Hinde, M. Dasgupta and M. Evers are warmly thanked for useful discussions. 
A.~W. is grateful to CEA/Saclay, IRFU/SPhN, where part of this work has been performed. 
The TDHF calculations were performed on the NCI National Facility
in Canberra, supported by the Commonwealth Government.
Support from ARC Discovery grants DP06644077 and DP110102858 is acknowledged.

\section*{References}
\bibliographystyle{epj}
\bibliography{biblio}

\end{document}